# Type II Supernovae as Probes of Cosmology

Dovi Poznanski[1,2], Eddie Baron[3], Stéphane Blondin[4], Joshua S. Bloom[1], Christopher B. D'Andrea[5], Massimo Della Valle[6], Luc Dessart[7], Richard S. Ellis[8], Avishay Gal-Yam[9], Ariel Goobar[10], Mario Hamuy[11], Malcolm Hicken[12], Daniel N. Kasen[13], Kevin L. Krisciunas[14], Douglas C. Leonard[15], Weidong Li[1], Mario Livio[16], Howie Marion[17], Thomas Matheson[18], James D. Neill[8], Ken'ichi Nomoto[19], Peter E. Nugent[2], Robert Quimby[8], Masao Sako[5], Mark Sullivan[20], Rollin C. Thomas[2], Massimo Turatto[21], Schuyler D. Van Dyk[22], W. Michael Wood-Vasey[23]

[1] UC Berkeley , [2] LBL, [3] University of Oklahoma, [4] ESO, [5] UPenn, [6] INAF-Capodimonte, [7] Laboratoire d'Astrophysique de Marseille, [8] Caltech, [9] Weizmann Institute, [10] Stockholm University, [11] Universidad de Chile, [12] CfA, [13] UC Santa Cruz, [14] Texas A&M, [15] San Diego State, [16] STScI, [17] University of Texas, [18] NOAO, [19] University of Tokyo, [20] Oxford University, [21] INAF-Catania, [22] SSC/Caltech, [23] University of Pittsburgh.

## ABSTRACT

**✦ Constraining the cosmological parameters and understanding Dark Energy have tremendous implications for the nature of the Universe and its physical laws.**

**✦ The pervasive limit of systematic uncertainties reached by cosmography based on Cepheids and Type Ia supernovae (SNe Ia) warrants a search for complementary approaches.**

**✦ Type II SNe have been shown to offer such a path. Their distances can be well constrained by luminosity-based or geometric methods. Competing, complementary, and concerted efforts are underway, to explore and exploit those objects that are extremely well matched to next generation facilities. Spectroscopic follow-up will be enabled by space-based and 20–40 meter class telescopes.**

**✦ Some systematic uncertainties of Type II SNe, such as reddening by dust and metallicity effects, are bound to be different from those of SNe Ia. Their stellar progenitors are known, promising better leverage on cosmic evolution. In addition, their rate – which closely tracks the ongoing star formation rate – is expected to rise significantly with look-back time, ensuring an adequate supply of distant examples.**

**✦ These data will competitively constrain the dark energy equation of state, allow the determination of the Hubble constant to 5%, and promote our understanding of the processes involved in the last dramatic phases of massive stellar evolution.**

# INTRODUCTION

The current preferred value for the Hubble Constant, is constrained to about 10%, as determined by the Hubble Key Project [1]. This value ultimately depends on the Cepheid period-luminosity relation in the LMC and other nearby galaxies – that remains remarkably uncertain (e.g., [2]). Such errors propagate directly up the distance scale and affect our ability to determine global cosmological parameters. For example, measurements of the total matter and baryonic content of the Universe via CMB or BAO experiments depend quadratically on $H_0$ [3,4], and in turn affect measurements of the Dark Energy density and its associated equation-of-state parameter [5,6]. Therefore, even a slight improvement in the knowledge of $H_0$ leads to a significant improvement in other cosmological parameters (see recent review in [7]).

Similar concerns plague Type Ia cosmography, as current statistics have virtually caught up with the inescapable barrier of systematic uncertainties. Despite a decade of use of SNe Ia as standard candles, we are still unclear of the physical nature of the SN Ia explosions. While exploring and constraining those uncertainties is one necessary path [8], complementary powerful methods should be developed.

Type II plateau SNe (SNe II-P) are the most frequent type of SN. Observationally they are defined by the presence of hydrogen in their spectra, and a “plateau” phase in their light curves. While SNe II-P are less luminous than SNe Ia, they are significantly more common, and their rate assuredly rises dramatically to a peak at redshifts 2-3 or beyond. The physics leading to their luminosity calibration is well understood, hence cosmic evolution is easier to study *a priori* (e.g., metallicity effects [9]). The progenitors of several objects have been found in archival images (e.g., [10,11]); all are red supergiants within a limited range of initial stellar masses. This suggests uniformity and perhaps only weak evolutionary trends. There are also recent indications that SNe II-P are quite uniform in the UV [12,13].

For more than 30 years, distance measurement methods using Type II SNe have been proposed and tested. Some rely on theoretical modeling, while others are mostly empirical, yet based on relatively well understood processes occurring in the thick hydrogen atmospheres that enshroud those SNe. Through recent methodological progress, combined with improved ground- and space-based facilities, SNe II-P offer the rich prospect of a great redshift baseline for cosmology and a valuable tracer of modes of star formation in host galaxies whose properties can be independently studied. In the following sections we describe the techniques, the current status in the field, and the promising prospects for different proposed missions.

# USING SPECTRAL MODELING TO DETERMINE $H_0$

Modern photospheric/spectroscopic methods use synthetic spectroscopy and spectral fitting to determine distances[14-24]. Model atmospheres are fit to spectroscopic and photometric observations, constraining the angular size of the SN, and thus the angular-diameter distance.

The good agreement in distances derived by competing groups to the same SNe (and agreement with Cepheid-based distances), using different models and methodology leads to confidence in the accuracy of the approach.

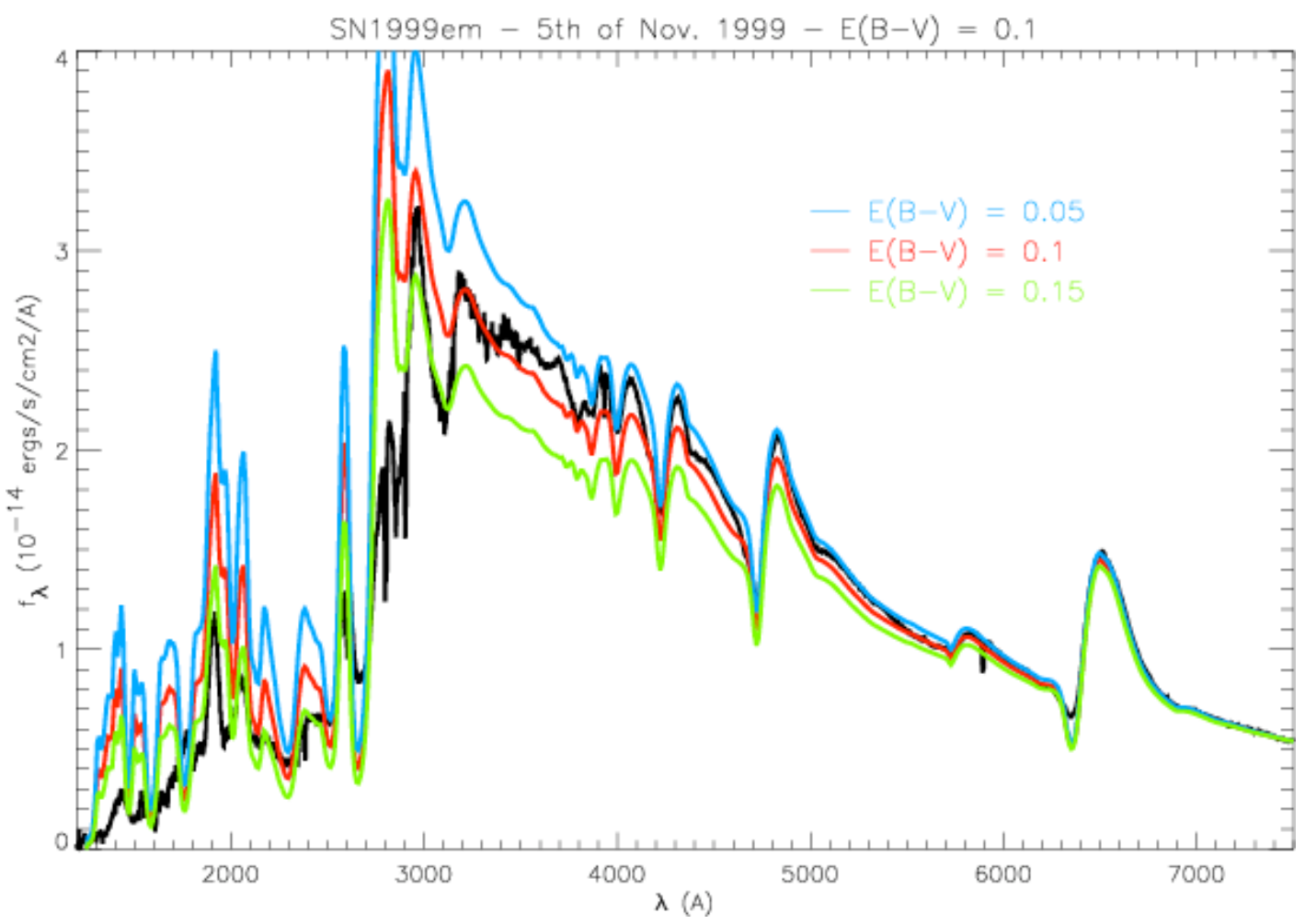

Figure 1: Model fit to the Type II-P SN 1999em, from [21]. Illustration of the effect of reddening modulations on the SED, and the power of synthetic spectroscopy in reconstructing the physical parameters of such a SN.

Such detailed calculations make metallicity effects largely irrelevant, as long as the photosphere is Hydrogen rich and optically thick. Model atmosphere calculations determine the reddening, while the ionization state of the ejecta is found independently through line fitting. Hence, continuum and line fluxes allow one to disentangle the effects of reddening from those of ionization/temperature. As a consequence, using SNe II-P to determine distances in the Universe is less subject to the systematic uncertainties seen when using SNe Ia.

The S/N required for these modeling-based methods inhibits rapid advance. Only a handful of nearby SNe are found every year, that are bright enough to be observed with current large telescopes. This also prohibits the use of this method at greater distances with current facilities. However, the proposed 20–40m class telescopes (TMT/ELT/GMT; hereafter - Giant Segmented Mirror Telescopes – GSMT) and the space-based JWST will push that limit significantly, allowing the analysis of more SNe, and more significantly, of more distant SNe in the Hubble flow. With a few nights on such a telescope, the Hubble constant could be constrained to better than 5%. This, combined with the Standardized Candle Method (see below) will allow the establishment of an independent distance ladder from the nearby Universe out to cosmological distances.

Another promising approach would be to better measure the period-luminosity relation for Cepheid, RR Lyrae, and Mira variables in the IR, where they should be less prone to dust and metallicity effects [25-29]. The same missions that we mention throughout this document, combined with parallaxes from the European astrometric mission – GAIA[1], will promote that goal through complementary means.

[1] To be launched in 2011.

# CONSTRAINING THE EQUATION OF STATE

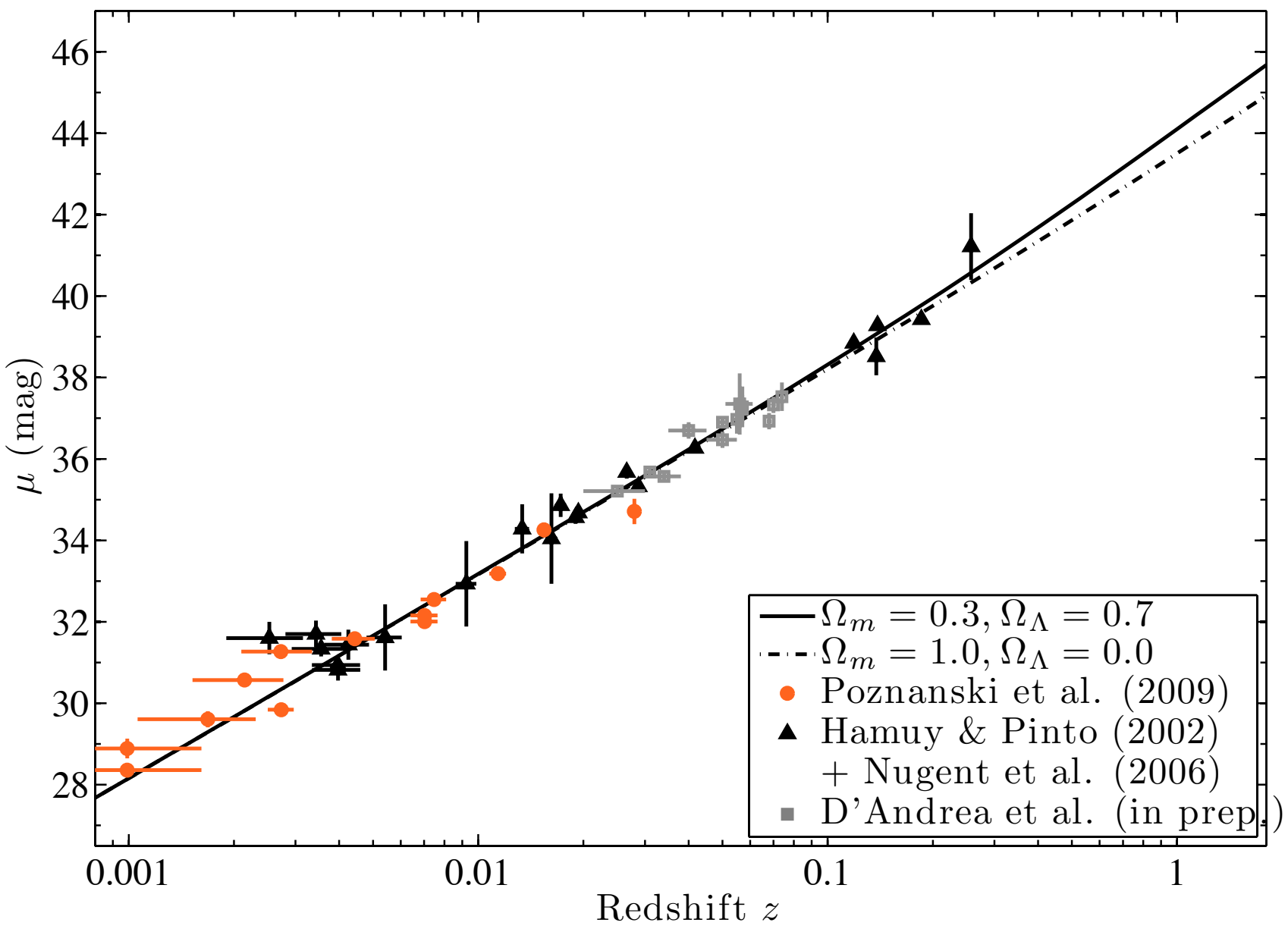

Figure 2: Current status of the Type II-P SN Hubble diagram with about 50 objects. The scatter is of the order of 10% in distance, only slightly worse than for SNe Ia. Most of those SNe are not in the Hubble flow, and have large peculiar velocity uncertainties. This will change with near-future samples (see text).

The standardized candle method first suggested by [30] provides an independent empirical way to achieve distances to SNe II-P, and is strongly anchored in simple physics. In more luminous SNe, the H-recombination front is maintained at higher velocities; the photosphere is farther out in radius. When a SN II-P is on the plateau phase of its light curve (which lasts for around 100 rest-frame days), there is a strong correlation between the velocity of the weak Fe II lines near 5000 Å – that trace the photospheric velocity – and the luminosity on the plateau. This empirical correlation was further studied [31-33] and the technique modified to simultaneously solve for the extinction correction (using the rest-frame V – I color on the plateau) and the Fe II velocity correction to arrive at a tight Hubble diagram with a systematic uncertainty of about 10% in distance, only slightly inferior to the current status for SNe Ia.

Given the small quantity of the data required – a few light curve points and a spectrum – this method is probably the only framework that could be applied cost-effectively at high redshifts. The number of SNe II-P for which distances have been derived is 38, with a similar number to be published soon by various groups [34-37].

With current uncertainties in the model (that should improve with upcoming greater samples) a handful of SNe at $z \sim 0.5$ are sufficient to independently measure the effects of acceleration to a $3\sigma$ confidence level. This is achievable with a few nights on current 8–10m telescopes.

According to our simulations a few nights on a GSMT would suffice to achieve the S/N necessary for about 20 SNe at $0.5 < z < 1$, and place substantial limits on the equation of state of dark energy. The exact limits depend on various details not yet determined, such as the capabilities of such an observatory, or on the progress we will see in the intervening years with lower-redshift SNe.

## DUST AND INTRINSIC PROPERTIES – CONSTRAINING SYSTEMATICS

One of the major systematic uncertainties in SN-Ia-based cosmology is the separation of intrinsic color scatter from extrinsic (both circumstellar and host-galaxy associated) dust [8]. There have been numerous measurements that point to a dust law that has significantly and consistently different wavelength dependence than the average measured in the Milky-Way [38-42]. In addition, there are indications that the same unexpected dust properties emerge when observing SNe II-P as well (e.g., [33]). It has recently been shown that at least some SNe II-P are very similar photometrically once dust extinction is accounted for [43,44].

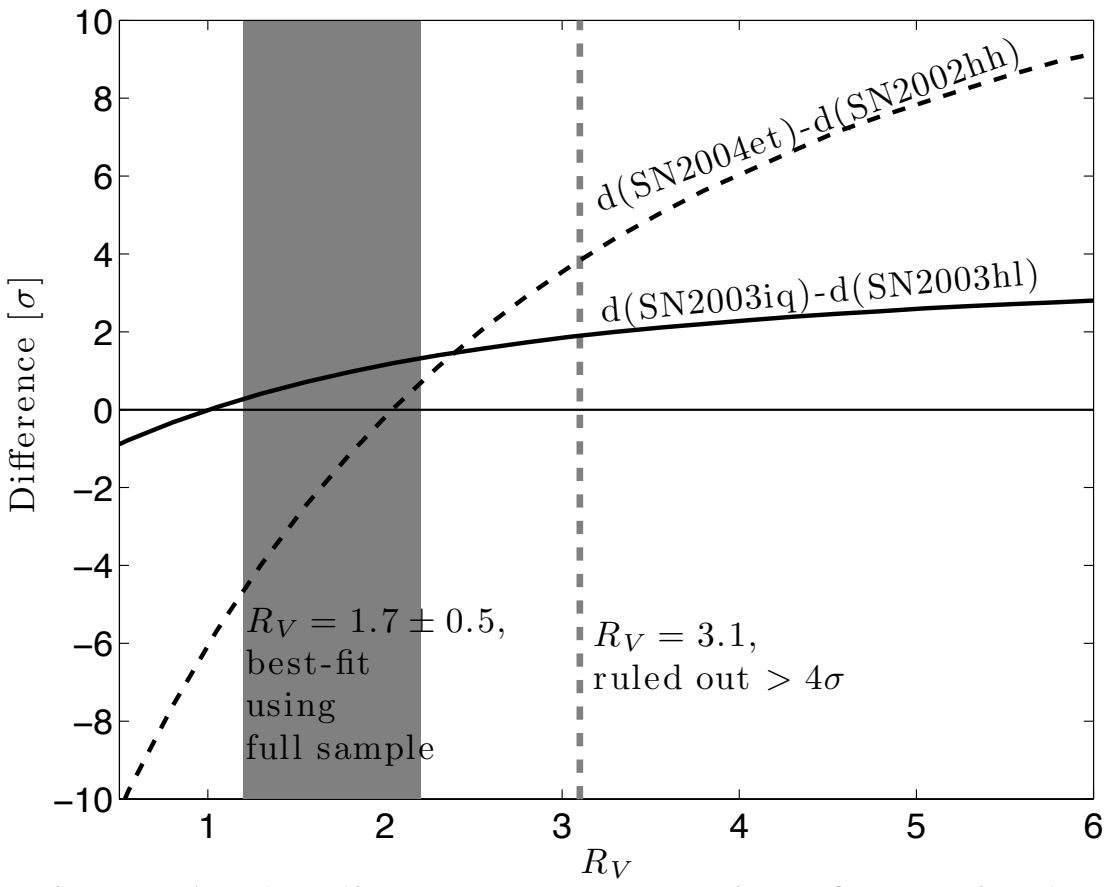

Figure 5: The distances to two pairs of SNe, in the same host galaxy constrain the extinction law to have small $R_V$ values. From [31].

Since color and reddening are inherently degenerate there is no trivial solution, unless one has a robust model for the underlying emission mechanism, as exists for SNe II-P. Any progress made in our understanding of the environments of one type of SN will fortify our constraints using the other.

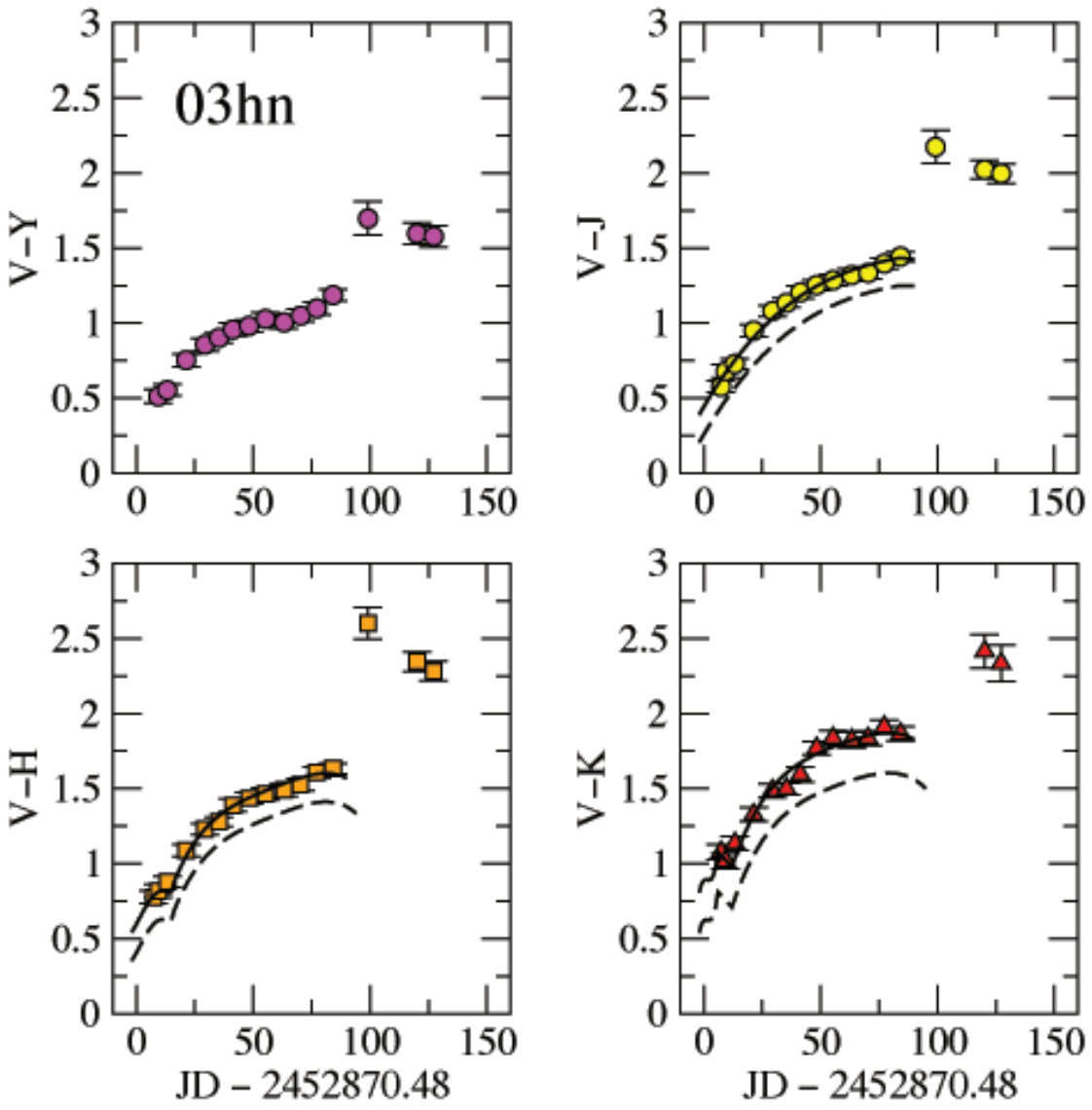

Figure 4: The colors of SN 2003hn, compared to SN 1999em, from [41]. The use of IR data allows one to constrain the contribution from dust extinction in the host galaxy, as opposed to intrinsic color.

Progress requires much larger samples of Type II SNe than currently available. Emerging surveys such as Pan-STARRS [45, SkyMapper [46], the Palomar Transient Factory (PTF; [47,48]), the La-Silla SN Search [49], will cover the low-$z$ range, to better constrain the different distance methods, and to probe the dust content—when combined with near-IR observations.

Future giant surveys such as the Dark Energy Survey (DES [50]), LSST [51] and SASIR[1] (see also [52]) will probe the medium-redshift range, in the optical to near-IR. The myriad of SNe II-P that these surveys will detect will exceed the follow-up capabilities that will be available. JDEM and instruments such as Hyper-SuprimeCam[2] will feed high-redshift studies. For those SNe, a GSMT and JWST will be crucial follow-up tools. SNe II-P seem to have a bright future mediated by budding and proposed missions. The existence of both empirical and theoretical methods allows for control of systematics. SNe II-P should become important complementary cosmological probes in the next decade.

[1] The Synoptic All-Sky Infrared Imaging survey. A pre-phase A concept for a new 6.5m telescope in Mexico, designed to repeatedly observe the IR sky simultaneously in 4 bands from 1–2.2 micron. First light is expected in 2016.

[2] An imager being built for the 8.2m SUBARU telescope, with a field of view of about 3 deg$^2$.

## REFERENCES

[1] Freedman, W. L. et al. 2001, *Final Results from the HST Key Project to Measure the Hubble Constant*, ApJ, 553, 47
[2] Romaniello, M. et al. 2005, *Chemical composition on Cepheids. I. P-L relation vs. iron abundance*, A&A, 429, L37
[3] Spergel, D. N. et al. 2007, *Three-Year WMAP Observations: Implications for Cosmology*, ApJS, 170, 377
[4] Eisenstein, D. J. et al. 2005, *Detection of the Baryon Acoustic Peak in the Large-Scale Correlation Function of SDSS Luminous Red Galaxies*, ApJ, 633, 560
[5] Garnavich, P. M. et al. 1998, *SN Limits on the Cosmic Equation of State*, ApJ, 509, 74
[6] Macri, L. M. et al. 2006, *Cepheid Distance to the Maser-Host Galaxy NGC 4258, Implications for $H_0$*, ApJ, 652, 1133
[7] Jackson, N. 2007, *The Hubble Constant*, Living Reviews in Relativity, 10, 4
[8] Howell, D. A. et al. 2009, *Astro2010 Decadal Survey Whitepaper:Type Ia SN Science 2010 – 2020*
[9] Baron, E. et al., 2003, *Determination of Primordial Metallicity and Mixing in the SN II-P 1993W*, ApJ, 586, 1199
[10] Van Dyk, S.D., Li, W., & Filippenko, A.V. 2003, *On the Progenitor of the Type II-P SN 2003gd in M74*, PASP, 115, 1289
[11] Smartt, S. J. et al. 2008, *The death of massive stars - I. Constraints on the progenitors of type II-P SNe*, ArXiv:0809.0403
[12] Gal-Yam, A. et al., 2008, *GALEX Spectroscopy of SN 2005ay Suggests UV Uniformity among SNe II-P*, ApJ, 685, L117
[13] Bufano, F. et al, 2009, ApJ submitted
[14] Baade, W. 1926, *Über eine Möglichkeit, die Pulsationstheorie der δ Cephei-Veränderlichen zu prüten*, Ast. Na., 228, 359
[15] Kirshner, R. P. & Kwan, J. 1975, *The envelopes of type II SNe*, *ApJ*, 197, 415
[16] Baron, E. et al. 1995, *SN 1993J: one year later*, Physics Reports, 256, 23
[17] Baron, E. et al. 1996, *Preliminary spectral analysis of SN 1994I, MNRAS*, 279, 799
[18] Mitchell, R. et al. 2002, *Detailed spectroscopic analysis of SN 1987A: The distance to the LMC using SEAM*, ApJ, 574, 293
[19] Baron, E. et al. 2004, *Type II-P SNe as cosmological probes: A ... distance to SN 1999em*, *ApJ*, 616, L91
[20] Dessart, L. & Hillier, D. J. 2005, *Quantitative spectroscopy of photospheric-phase type II SNe*, A&A, 437, 667
[21] Dessart, L. & Hillier, D. J. 2006, *Quantitative spectroscopic analysis of and distance to SN1999em.* A&A, 447, 691
[22] Baron, E. et al. 2007, *Reddening, abundances, and line formation in SNe II*, ApJ, 662, 1148
[23] Dessart, L. & Hillier, D. J. 2008, *Time-dependent Effects in Photospheric-Phase Type II SN Spectra*, MNRAS, 383, 57
[24] Dessart, L. et al. 2008, *Using Quantitative Spectroscopic Analysis to Determine the Properties and Distances of Type II-P SNe: SN 2005cs and SN 2006bp*, ApJ, 675, 644
[25] Bono, G., 2003, *Cosmic Distances: Current Odds and Future Perspectives*, ASPC, 291, 45
[26] Sollima, A., Cacciari, C. & Valenti, E., 2006, *The RR Lyrae period-K-luminosity relation for globular clusters: an observational approach*, MNRAS, 372, 1675
[27] Whitelock, P. A., Feast, M. W. & van Leeuwen, F., 2008, *AGB variables and the Mira P-L relation*, MNRAS, 386, 313
[28] Feast, M. W., 2008, *Galactic and Extragalactic Distance Scales: The Variable Star Project*, arXiv/0806.3019
[29] Tammann, G. A. et al., 2008, *Distances from RR Lyrae, the Tip of the RGB, and Classical Cepheids*, ApJ, 679, 52
[30] Hamuy, M. & Pinto, P. A., 2002, *Type II SNe as Standardized Candles*, ApJ, 566, L63
[31] Hamuy, M., 2005, *The Standard Candle Method for Type II SNe and $H_0$,* in IAU 192, 535
[32] Nugent, P. et al., 2006, *Toward a Cosmological Hubble Diagram for Type II-P SNe*, ApJ, 645, 841
[33] Poznanski, D. et al., 2008, *Improved Standardization of Type II-P SNe: Application to an Expanded Sample*, ArXiv/0810.4923, ApJ accepted
[34] Olivares, F., 2008, *The Standard Candle Method for Type II-Plateau SNe*, Master Thesis
[35] D'Andrea, C. et al., 2009, *Type II-P SNe from SDSS II SN Survey*, BAAS, 213, 425.14
[36] Emilio Enriquez, J. et al., 2009, *Distances to Type II-P SNe from the Caltech Core-Collapse Project*, BAAS, 213,490.07
[37] Poznanski, D. et al., 2009, *in preparation*
[38] Elias-Rosa, N. et al., 2006, *Anomalous extinction behaviour towards the Type Ia SN 2003cg*, MNRAS, 369, 1880
[39] Krisciunas, K. et al., 2007, *The Type Ia SN 2004S, a Clone of SN 2001el, and the Optimal Photometric Bands for Extinction Estimation*, AJ, 133, 58
[40] Elias-Rosa, N. et al., 2008, *SN 2002cv: a heavily obscured Type Ia SN*, MNRAS, 384, 107
[41] Nobili, S. & Goobar, A., 2008, *The colour-lightcurve shape relation of type Ia SNe and the reddening law,* A&A, 487, 19
[42] Wang, X. et al., 2008, *Optical and IR Observations of the ... Type Ia SN 2006X in M100*, ApJ, 675, 626
[43] Krisciunas, K. et al., 2009, *Do the Colors of SNe II-P Allow Determination of Host Extinction?*, AJ, 137, 34
[44] Olivares E. F., & Hamuy, M. 2009, *in preparation*
[45] Kaiser, N. et al., 2002, Pan-STARRS: A Large Synoptic Survey Telescope Array, SPIE Proc., 4836, 154
[46] Keller, S. C. et al., 2007, *The SkyMapper Telescope and The Southern Sky Survey*, PASA, 24, 1
[47] Rau, A. et al., 2009, *in preparation*
[48] Poznanski, D. et al., 2009, *The SCM For Type II-P SNe - New Sample And PTF Perspective*, BAAS 213, 469.09
[49] Baltay, C. et al., 2007, *The QUEST Large Area CCD Camera*, PASP, 119, 1278
[50] The Dark Energy Survey Collaboration, 2005, *The Dark Energy Survey*, ArXiv: 0510346
[51] Ivezic, Z. et al., 2008, *LSST: from Science Drivers to Reference Design and Anticipated Data Products*, ArXiv/0805.2366
[52] Bloom, J. S. et al., 2009, *Astro2010 Whitepaper: Coordinated Science in the Gravitational and EM Skies*